\documentclass[amsmath,amssymb,aps,prb,superscriptaddress,longbibliography,citeautoscript]{revtex4-1}
\usepackage[margin=1in]{geometry}			
\usepackage{graphicx,epsfig}
\usepackage{tikz}
\usepackage{xcolor}
\usepackage{hyperref}
\hypersetup{
    colorlinks = true,
}
\usepackage{comment}
\usepackage{bibunits}
\usepackage{lineno}

\begin{document}
\author{Yu Liu}
\thanks{These authors contributed equally.}
\affiliation{Department of Chemistry and Chemical Biology, Harvard University, Cambridge, Massachusetts, 02138, USA}
\affiliation{Department of Physics, Harvard University, Cambridge, Massachusetts, 02138, USA}
\affiliation{Harvard-MIT Center for Ultracold Atoms, Cambridge, Massachusetts, 02138, USA}

\author{Ming-Guang Hu}
\thanks{These authors contributed equally.}
\affiliation{Department of Chemistry and Chemical Biology, Harvard University, Cambridge, Massachusetts, 02138, USA}
\affiliation{Harvard-MIT Center for Ultracold Atoms, Cambridge, Massachusetts, 02138, USA}

\author{Matthew A. Nichols}
\thanks{}
\affiliation{Department of Chemistry and Chemical Biology, Harvard University, Cambridge, Massachusetts, 02138, USA}
\affiliation{Harvard-MIT Center for Ultracold Atoms, Cambridge, Massachusetts, 02138, USA}

\author{David D. Grimes}
\thanks{}
\affiliation{Department of Chemistry and Chemical Biology, Harvard University, Cambridge, Massachusetts, 02138, USA}
\affiliation{Harvard-MIT Center for Ultracold Atoms, Cambridge, Massachusetts, 02138, USA}

\author{Tijs Karman}
\thanks{}
\affiliation{ITAMP, Harvard-Smithsonian Center for Astrophysics, Cambridge, Massachusetts, 02138, USA}

\author{Hua Guo}
\thanks{}
\affiliation{Department of Chemistry and Chemical Biology, University of New Mexico, Albuquerque, New Mexico, 817131, USA}

\author{Kang-Kuen Ni}
\thanks{}
\affiliation{Department of Chemistry and Chemical Biology, Harvard University, Cambridge, Massachusetts, 02138, USA}
\affiliation{Department of Physics, Harvard University, Cambridge, Massachusetts, 02138, USA}
\affiliation{Harvard-MIT Center for Ultracold Atoms, Cambridge, Massachusetts, 02138, USA}

\title{{\Large Steering ultracold reactions through long-lived transient intermediates}}
\date{\today}
\begin{abstract}

\begin{large}
Controlling the pathways and outcomes of reactions is a broadly pursued goal in chemistry. In gas phase reactions, this is typically achieved by manipulating the properties of the reactants, including their translational energy \cite{qiu2006observation}, orientation \cite{pan2017direct}, and internal quantum state \cite{puri2019reaction}.
In contrast, here we influence the pathway of a reaction via its intermediate complex, which is generally too short-lived to be affected by external processes.
In particular, the ultracold preparation of potassium-rubidium (KRb) reactants leads to a long-lived intermediate complex (K$_2$Rb$_2^*$), which allows us to steer the reaction away from its nominal ground-state pathway onto a newly identified excited-state pathway using a laser source at 1064~nm, a wavelength commonly used to confine ultracold molecules.
Furthermore, by monitoring the change in the complex population after the sudden removal of the excitation light, we directly measure the lifetime of the complex to be $360 \pm 30$ ns, in agreement with our calculations based on the Rice-Ramsperger-Kassel-Marcus (RRKM) statistical theory \cite{levine2009molecular}.
Our results shed light on the origin of the two-body loss widely observed in ultracold molecule experiments \cite{ospelkaus2010quantum,takekoshi2014ultracold,ye2018collisions, gregory2019sticky, park2015ultracold}.
Additionally, the long complex lifetime, coupled with the observed photo-excitation pathway, opens up the possibility to spectroscopically probe the structure of the complex with high resolution, thus elucidating the reaction dynamics.
\end{large}

\end{abstract}
\maketitle
\large

Chemical reactions between quantum-state-controlled molecules at ultralow temperatures display unique characteristics, 
including a strong dependence of the reaction rate on long-range interaction potentials \cite{ospelkaus2010quantum, ni2010dipolar, de2011controlling, puri2019reaction, hall2012millikelvin}, scattering resonances that are extremely narrow in energy \cite{mcdonald2016photodissociation, klein2017directly, yang2019observation}, and reactive pathways dominated by single scattering channels \cite{rui2017controlled,hoffmann2018reaction}. Even the elusive transient intermediate complex was directly observed in the ultracold bimolecular reaction of KRb molecules \cite{hu2019direct}, where the  preparation of rovibronic ground-state molecules at sub-microkelvin temperatures energetically minimized the number of product channels available for the complex to dissociate, resulting in its long lifetime. It is natural, then, to inquire about the exact lifetime of this complex. Additionally, one may consider the consequences of this long lifetime, as well as the possibility to control the reaction at this stage.

The role of long-lived complexes has been discussed extensively in the context of ultracold collisions between rovibronic ground-state bialkali molecules (AB) where the atom-exchange pathway to reaction products (AB + AB $\rightarrow$ A$_2$ + B$_2$) is energetically forbidden, but two-body loss of molecules was experimentally detected \cite{takekoshi2014ultracold, ye2018collisions,  gregory2019sticky, park2015ultracold}.
Such losses limit the phase space densities achievable in these systems, and present a major obstacle to realizing novel quantum phases with long-range, dipolar interactions in bulk molecular samples \cite{santos2000bose, buchler2007strongly, levinsen2011topological, baranov2012condensed}.
It was proposed that a long-lived complex (A$_2$B$_2^*$) allows time for a third body (a photon~\cite{christianen2019photoinduced} or a molecule~\cite{mayle2013scattering}) to interact with the complex, leading to the observed loss.
However, the complex lifetimes calculated in these proposals are widely different \cite{mayle2013scattering,christianen2019quasiclassical}, which stems from difficulties in estimating the density of states of the complex and uncertainties in whether nuclei maintain their spins during reactions. The loss mechanisms therefore remain an open question.

In this article, by studying reactive collisions between $^{40}$K$^{87}$Rb molecules in the presence of varying intensities of 1064~nm light, we observe significant photo-excitation of the complex, manifested as a reduction in both the complex and the product populations. We determine the first-order rate constant for the excitation to be $0.42 \pm 0.09 ~\mu\textrm{s}^{-1}/(\textrm{kW/cm}^2)$, in agreement with our theoretical prediction.
Exploiting this effect, we use light to induce complex loss and, after an abrupt shut-off of the light, monitor the time evolution of the complex population as it reaches a steady state to extract the complex lifetime. 
Our measurement agrees well with the calculated value based on RRKM theory using full-dimensional \textit{ab initio} potential energy surfaces (PESs).

\section{Ultracold reactions in the presence of light}

\begin{figure}[t!]
\centering
\includegraphics[width = 0.9\textwidth]{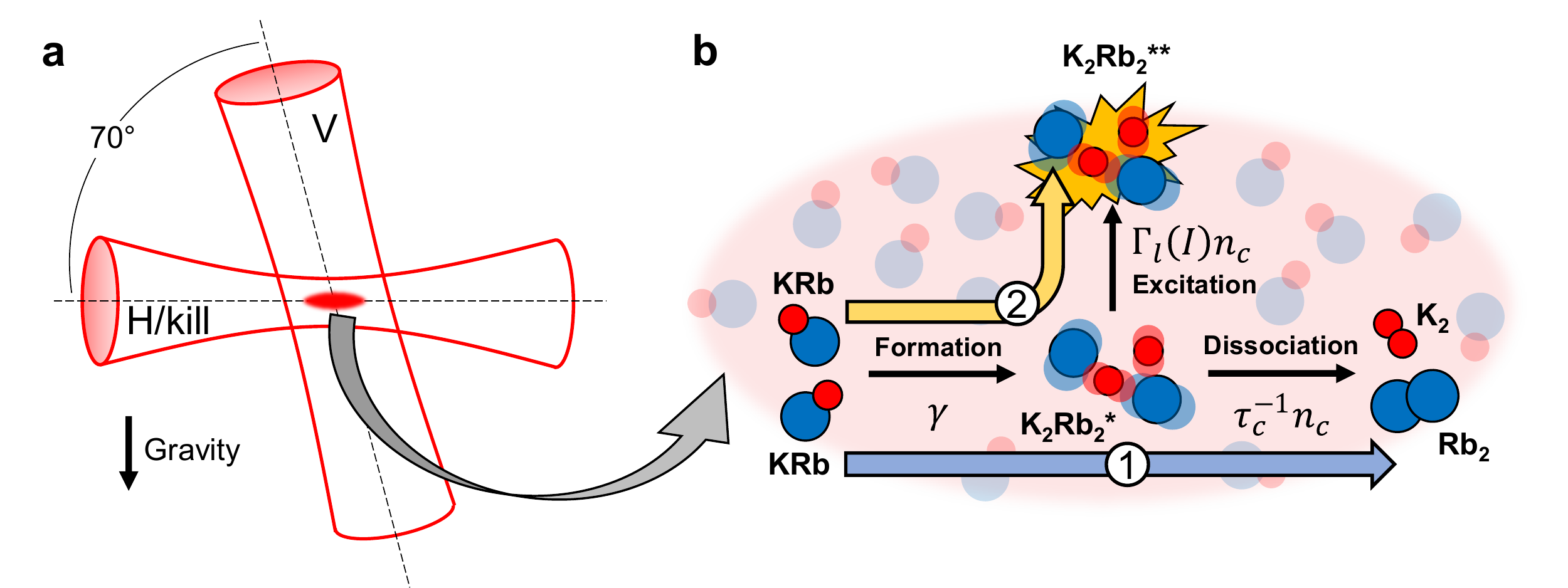}
\caption{\textbf{Ultracold reactions in an optical dipole trap}. \textbf{a,} Optical trapping of a gas of KRb molecules. ``H'' and ``V'' are Gaussian beams with $1/e^2$ diameters of 60 and 200 $\mu$m, respectively. The ``kill'' beam, introduced for the purpose of complex lifetime measurement, shares the same optical path with ``H'', but has independent timing control. \textbf{b,} Two pathways for bimolecular KRb reactions. Both pathways involve the formation of a long-lived intermediate complex K$_2$Rb$_2^*$. \textcircled{1} indicates the ground state pathway KRb + KRb $\rightarrow$ K$_2$Rb$_2^*$ $\rightarrow$ K$_2$ + Rb$_2$, which is the only pathway in the absence of ODT light; \textcircled{2} indicates the excited state pathway KRb + KRb $\rightarrow$ K$_2$Rb$_2^*$ $\rightarrow$ K$_2$Rb$_2^{**}$, which is dominant at high ODT intensity.
The rates for K$_2$Rb$_2^*$ formation, dissociation, and excitation are labeled as $\gamma$, $\tau_c^{-1}n_c$, and $\Gamma_l(I)n_c$, respectively.}
\label{figSchematic}
\end{figure}

Each experiment begins with a gas of $\sim5000$ rovibronic ground state KRb molecules prepared at a temperature of 500~nK and an average number density of $3.5 \times 10^{11} ~\textrm{cm}^{-3}$ inside a crossed optical dipole trap (ODT) formed from two 1064~nm laser beams (``H'' and ``V''), as illustrated in Fig. \ref{figSchematic}(a). Details of the apparatus regarding the production and detection of the gas were reported previously ~\cite{liu2020probing}. As soon as the KRb sample is created inside the ODT, its density, $n_r$, decays via two-body loss that can be characterized by $n_r(t) = n_r(t = 0)(1 + t/t_{1/2})^{-1}$,
where $t_{1/2}$ denotes the half-life, which is empirically measured to be $250\pm30$ ms. In a previous study \cite{hu2019direct} it was determined that, in the absence of ODT light, this population decay occurs as a result of the exothermic bimolecular exchange reaction, 
\begin{equation} \label{KRb+KRb}
^{40}\textrm{K}^{87}\textrm{Rb}\,+\,^{40}\textrm{K}^{87}\textrm{Rb} \rightarrow \textrm{K}_{2}\textrm{Rb}_{2}^* \rightarrow \textrm{K}_2\,+\,\textrm{Rb}_2\,+\,10~\textrm{cm}^{-1},
\end{equation}
where $^*$ denotes the transient intermediate complex (Fig. \ref{figSchematic}(b)). Here, we study this reaction in the presence of the ODT, and watch for possible effects of the 1064~nm light.

\begin{figure}[ht!]
\centering
\includegraphics[width = 0.75\textwidth]{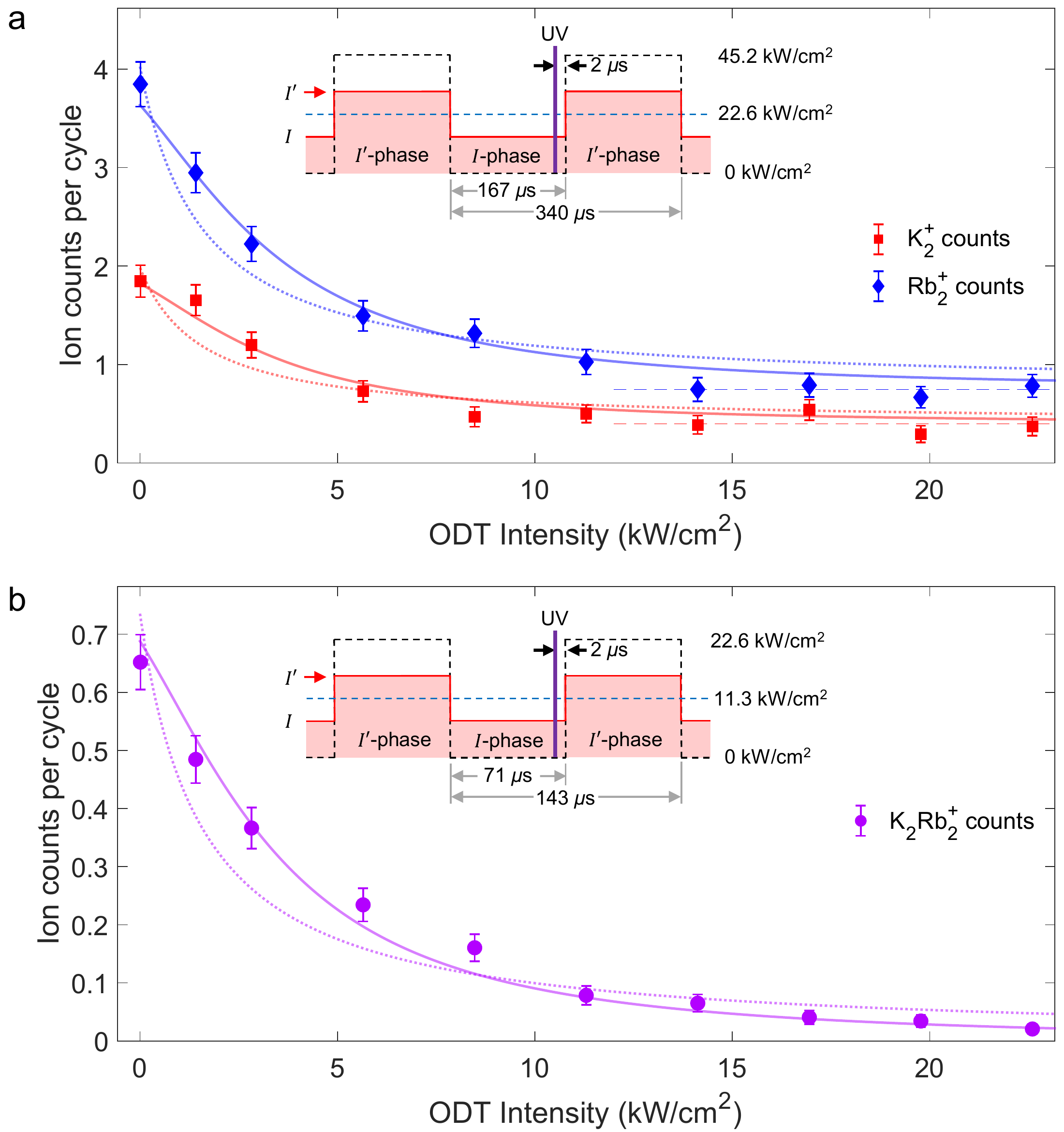}
\caption{\textbf{Steering the reaction pathway via the intermediate complex using 1064~nm light.} 
\textbf{a}, Steady-state K$_2^+$ (red squares) and Rb$_2^+$ (blue diamonds) ion counts at various ODT light intensities, normalized by the number of experimental cycles ($\sim 80$ for each data point). The error bars represent shot noise.
The dashed lines indicate the offset levels $C_{\textrm{K}_2^+}$ and $C_{\textrm{Rb}_2^+}$. 
\textbf{b}, Steady-state K$_2$Rb$_2^+$ (purple circles) ion counts at various ODT light intensities, normalized by the number of experimental cycles ($ \sim 290 $ for each data point).
The solid lines are simultaneous fits to the data in \textbf{a} and \textbf{b} using Eq. \ref{IonSteadyState} with $B_{1}$ and $B_{2}$ as free parameters, and the dotted lines are fits with $B_{2}$ set to zero and only $B_{1}$ as a free parameter.
The fits are weighted by the inverse of the ion counts.
(Insets) Timing schemes for the ODT (red) and the pulsed UV ionization laser (purple) used for the product and complex measurements. The blue dashed line represents the time-averaged intensity level, while the black dashed line represents the ODT intensity envelope for full depth modulation.}
\label{figODTEffect}
\end{figure}

We first probe for changes in the amount of reaction products (K$_2$ and Rb$_2$) formed while the KRb cloud is exposed to varying intensities of ODT light. A simple change in the overall intensity of the ODT, however, would modify the confining potential felt by the KRb molecules and therefore the temperature and density of the gas, resulting in changes in the rate of reaction. To circumvent this issue, we apply a fast square wave modulation to the ODT intensity (Fig. \ref{figODTEffect}(a) inset). This creates two phases with variable instantaneous intensity levels, $I$ and $I'$, while keeping the time-averaged intensity, $ (I + I')/2$, constant. As long as the modulation frequency, $f_{\textrm{mod}}$, is much higher than the trapping frequencies of the gas, $f_{\textrm{trap}}$, the molecular motions are unaffected, and therefore the density and temperature of the gas remain unchanged. We fulfill this requirement by choosing $f_{\textrm{mod}} = 3$ kHz, where $f_{\textrm{trap}} \leq 0.4$ kHz along all three axes of the trap.

As the reaction proceeds, we sample the product population using ion time-of-flight mass spectrometry (TOF-MS), the details of which were described previously \cite{hu2019direct,liu2020probing}. In brief, the region surrounding the KRb cloud is exposed to a pulsed UV ionization laser with a wavelength of 305~nm and a pulse duration of 7 ns. Photoionized products are then accelerated onto a time and position sensitive ion detector and counted. The total ion counts associated with each species serves as a proxy for its steady-state population at the instances of the UV pulses. The repetition of the UV pulses is synchronized to the ODT intensity modulation such that the reaction is always probed towards the end of the $I$-phase, allowing the product distribution to settle into steady state after the previous $I'$-phase.

We accumulated ion data at 10 different values of $I$ between 0 and 22.6 $\textrm{kW}/\textrm{cm}^2$, with a fixed time-averaged intensity of 22.6 $\textrm{kW}/\textrm{cm}^2$. The counts of K$_2^+$ and Rb$_2^+$, plotted in Fig. \ref{figODTEffect}(a), decrease monotonically with increasing values of $I$, indicating a reduction in the amount of products formed as the ODT becomes more intense. We rule out electronic excitation of the products by the 1064~nm light as the cause of this reduction, since the light is far-detuned from any relevant molecular transitions in K$_2$~\cite{vadla2006comparison} and Rb$_2$~\cite{edvardsson2003calculation}. Additionally, we observe that the ion counts plateau at high ODT intensities to a value $\sim 25\%$ of the maximum signal, an effect which persists up to at least 45.2 $\textrm{kW}/\textrm{cm}^2$, as we have independently verified (see methods).

Because products are formed from the dissociation of the K$_2$Rb$_2^*$ complexes, it is natural to question how the complexes are affected by the ODT light. To this end, we probe the complex population at different ODT intensities using an experimental protocol similar to that used for the products (Fig. \ref{figODTEffect}(b) inset). Here, the UV ionization laser is tuned to 354.77~nm, a wavelength that results in the photoionization of K$_2$Rb$_2^*$ into K$_2$Rb$_2^+$ \cite{hu2019direct}. The intensity of the ODT is then modulated at $f_{\textrm{mod}} = 7$ kHz with a time-averaged value of 11.3 $\textrm{kW}/\textrm{cm}^2$. K$_2$Rb$_2^+$ ion counts are accumulated at 10 different values of $I$ between 0 and 22.6 $\textrm{kW}/\textrm{cm}^2$, as shown in Fig. \ref{figODTEffect}(b).

\begin{figure}[t!]
\centering
\includegraphics[width = 1.0\textwidth]{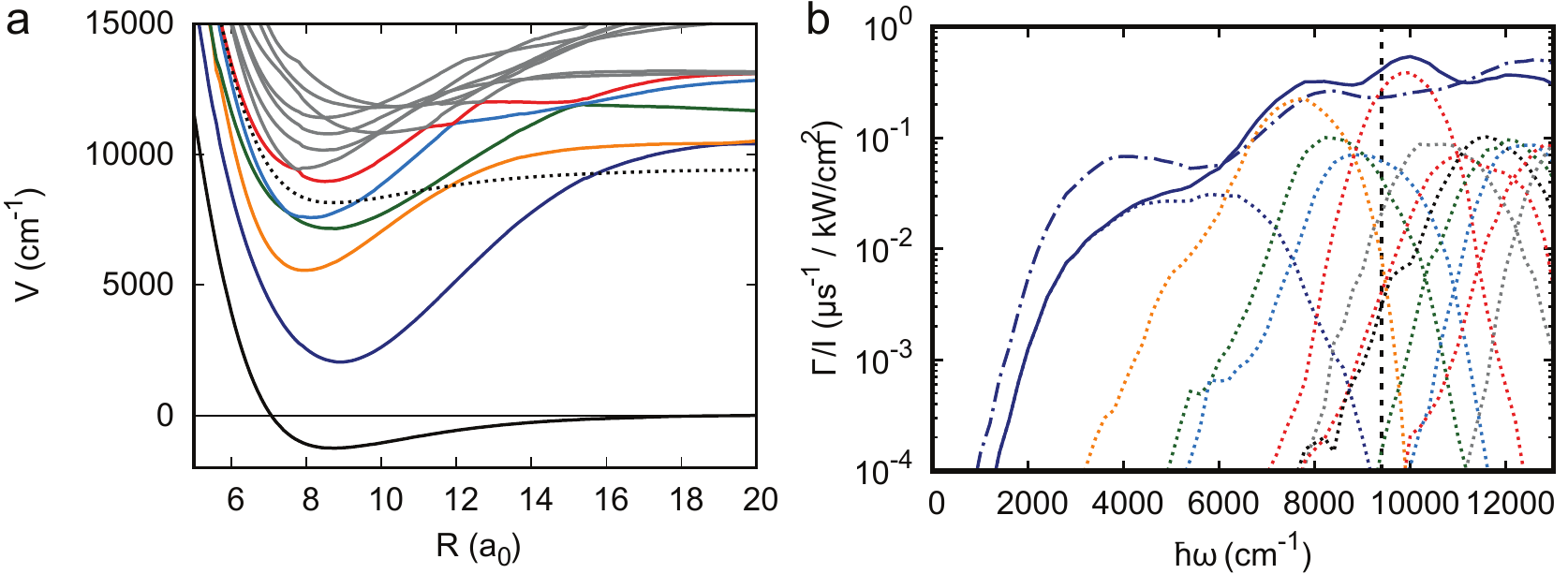}
\caption{\textbf{Energies and rates of the electronic excitations of the complex}. \textbf{a}, Potential energy curves for the ground state and low-lying excited states of KRb+KRb as a function of the distance between the molecules, $R$, in units of the Bohr radius, $a_0$, for fixed orientation and monomer bond lengths.
The black dotted curve shows the ground-state potential shifted vertically by the energy of a 1064~nm photon.
This crosses various excited state potentials, indicating these are accessible through one-photon excitation of the complex.
\textbf{b},
Calculated excitation rates as a function of the excitation photon energy,
with the 1064~nm ODT marked by the vertical dashed line.
Dotted lines show the contribution of various adiabatic electronic states, color-coded accordingly with the potential curves in \textbf{a}.
The blue solid line shows the corresponding total excitation rate,
whereas the dashed-dotted line shows the excitation rate obtained assuming the total dipole moment is equally distributed over all excited states.
The difference between the two serves as an indication of the uncertainty in the calculated transition dipole moments.
}
\label{figTheory}
\end{figure}

Similar to the products, the complex population also decreases with increasing intensity of the 1064 nm light. This observation points to photo-induced complex loss, as postulated in Ref. \cite{christianen2019photoinduced}. To verify this hypothesis, we computed the energies and rates of electronic excitations of K$_2$Rb$_2$. The results show that multiple excited states can be reached by the absorption of a single 1064 nm photon, with an appreciable transition rate at most of the ODT intensities explored here (Fig. \ref{figTheory}). This indicates that, in addition to the previously observed ground state pathway, there is a competing excited pathway (Fig. \ref{figSchematic}(b)),
\begin{equation}
    \textrm{KRb} + \textrm{KRb} \rightarrow \textrm{K}_2\textrm{Rb}_2^* \xrightarrow{h\nu} \textrm{K}_2\textrm{Rb}_2^{**}.
\end{equation}
The photo-excited complex, K$_2$Rb$_2^{**}$, is a potential source of the offset observed at high ODT intensities in the product data of Fig. \ref{figODTEffect}(a). That is, it may be possible for a small fraction of the K$_2$Rb$_2^{**}$ population to spontaneously decay into product forming channels on the ground potential surface, which are subsequently ionized and then detected.

We include both pathways to model the data in Fig. \ref{figODTEffect} using the rate equation
\begin{equation} \label{complexRateEq}
    \dot{n}_c(t) = \gamma - \tau_c^{-1} n_c(t) - \Gamma_l(I) n_c(t),
\end{equation}
where $n_c$ is the complex density, $\gamma = -\dot{n}_r/2$ is the rate of KRb + KRb collisions, $\tau_c$ is the complex lifetime, and $\Gamma_l$ is the intensity-dependent photo-excitation rate. In steady state, the formation of the complex from reactants is balanced by its dissociation into products, and its photo-excitation, so that $\dot{n}_c = 0$. In this case, the complex density is given by 
\begin{equation} \label{complexDensitySteadyState}
    n_c = \frac{\gamma}{\tau_c^{-1} + \Gamma_l(I)}.
\end{equation}
We model the excitation rate as $\Gamma_l = \beta_1 I + \beta_2 I^2$, where the linear term represents single-photon excitation of the complex, and the quadratic term represents possible second-order contributions. 
Equation \ref{complexDensitySteadyState} then becomes $n_c = \gamma\tau_c\left(1 + B_1 I + B_2 I^2 \right)^{-1}$, where $B_{1,2} = \beta_{1,2}~\tau_c $. 
Since both the product and the complex ion counts are proportional to the steady-state density of the complex, their dependence on the 1064 nm light intensity can be modeled as
\begin{equation} \label{IonSteadyState}
N_s = \frac{A_s}{1 + B_{1}I + B_{2}I^2} + C_\textrm{s},
\end{equation}
where $s = \textrm{K}_2^+, \textrm{Rb}_2^+, \textrm{or} ~ \textrm{K}_2\textrm{Rb}_2^+$, and $C_s$ is introduced to account for any offsets in the data.
We apply Eq. \ref{IonSteadyState} to simultaneously fit the product and complex ion data, as shown in Fig. \ref{figODTEffect}(a) and (b). For the product data, $C_s$ is fixed to be the average value of the respective set of four data points at the highest ODT intensities, while for the complex data, $C_s$ is set to zero. We allow the amplitudes $A_s$ to be independent free parameters, whereas $B_1$ and $B_2$ are shared parameters across the data sets. The fit adequately captures the data in all three sets, and yields $B_1 = 0.15\pm0.03 ~ \textrm{cm}^2/\textrm{kW}$ and $B_2 = 0.050\pm0.004 \left(\textrm{cm}^2/\textrm{kW}\right)^2$, with a mean-squared-error (MSE) of 1.2. These values for $B_{1,2}$ indicate that, at a typical ODT intensity of $11.3 ~\textrm{kW/cm}^2$, the reaction is $\sim 8$ times as likely to proceed via the excited-state pathway as the ground-state pathway. We thus find that, by controlling the optical intensity of the 1064~nm light, efficient steering of the reaction pathway can be achieved.

We also fit the three data sets in Fig. \ref{figODTEffect} with $B_2$ set to zero, representing an excitation rate model that is purely linear in the intensity. In this case we find that $B_1 = 0.64 \pm 0.05 ~\textrm{cm}^2/\textrm{kW}$, but that the fit yields a large MSE of 4.0, and systematically underestimates the lower intensity data points while overestimating the higher ones. This indicates that the linear rate model is insufficient to fully describe the observed photo-excitation of the complex.

\begin{figure}[t!]
\centering
\includegraphics[width = 1.0\textwidth]{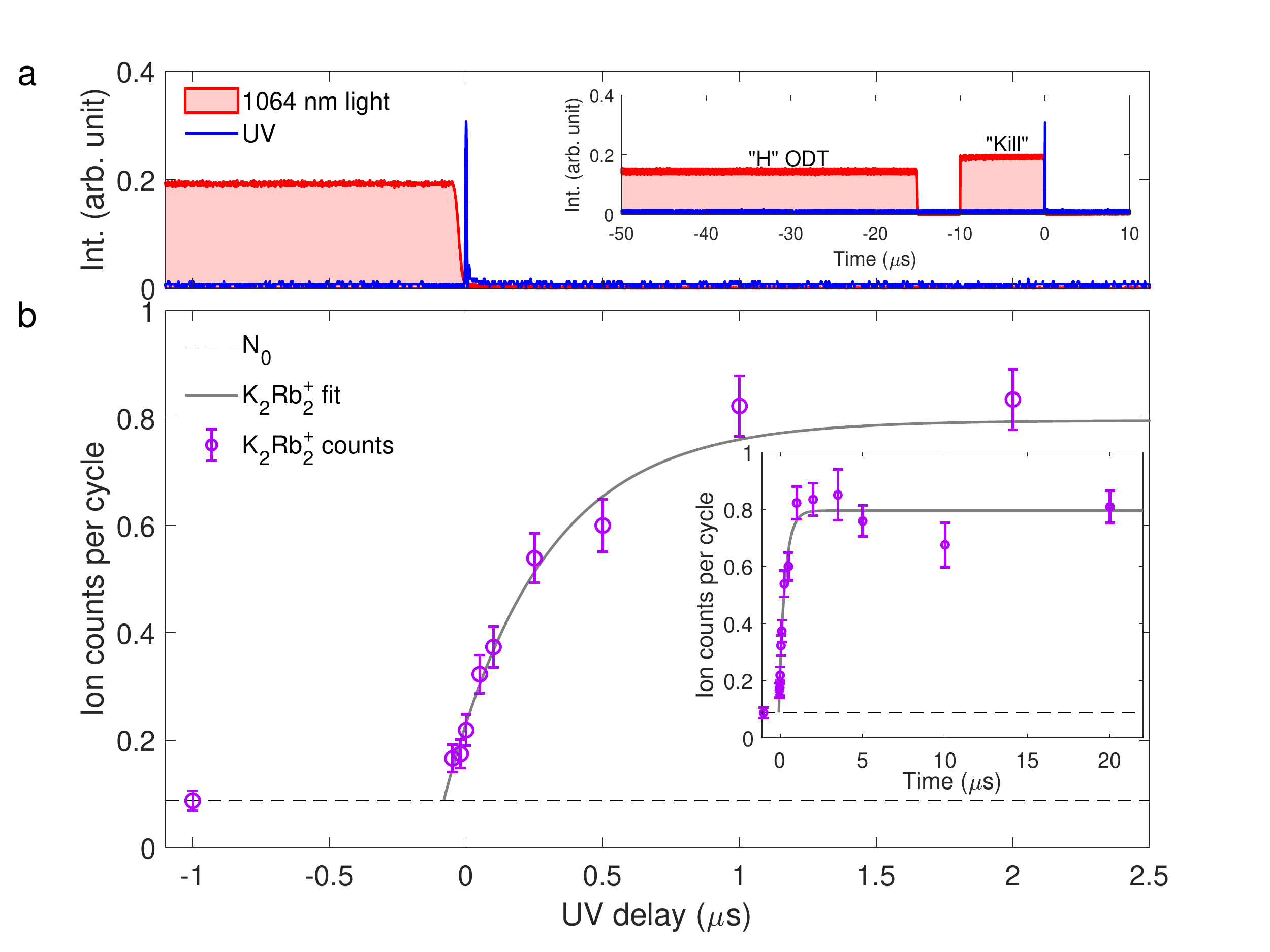}
\caption{\textbf{Lifetime of the intermediate complex.} \textbf{a,}  Measured time traces of the 1064~nm light (red) and the UV ionization pulse (blue) intensities. The UV pulses probe the complex population around the off edge of the ``kill'' pulse, which provides a zero-of-time for the measurements. (Inset) The same traces displayed over a wider time window showing the relative timing between the ``H'' and ``kill'' beams. The ``V'' beam, not shown here, shares the same timing as the ``H'' beam.
With the addition of the ``kill'' beam, the ``H'' beam power is adjusted such that the time-averaged intensity of the ODT is maintained at 11.3 kW/cm$^2$. \textbf{b,} K$_2$Rb$_2^+$ ion counts (purple circles) measured at different UV delays with respect to the off edge of the ``kill'' pulse. The dashed line indicates the baseline from which the ion counts grow after the light is turned off. The level for this baseline is given by the ion counts at $-1 \mu$s, $N_0$. The solid line shows a fit to the entire data set (see inset) using the function $A\left[1 - (1-N_0/A)\textrm{e}^{-(t - t_0)/\tau_c}\right]$, which is derived from Eq. \ref{complexGrowth}. The fit yields a complex lifetime $ \tau_c = 360 \pm 30 $ ns, and a timing offset $ t_0 = -82 \pm 7 $ ns. This offset accounts for a systematic uncertainty in the relative timing between the 1064~nm light and the UV. The MSE for the fit is 0.85.
}
\label{figComplexLifetime}
\end{figure}

\section{Measuring the complex lifetime}

The lifetime of the intermediate complex has been measured for a variety of chemical systems \cite{sato2001photodissociation, noll1998bimolecular, scherer1990real}, where the use of ultrafast lasers and molecular beam techniques establishes a well-defined zero-of-time for all individual reaction events, allowing the process of complex formation and dissociation to be monitored in real time. In our experiment, the bulk nature of the ultracold KRb sample makes establishing a zero-of-time challenging, as individual reactions occur stochastically. Fortunately, the observed optical excitation of the complex makes it possible to use light to set a zero-of-time. Specifically, we first strongly reduce the complex population by exposing the KRb cloud to intense 1064 nm light, and then monitor its growth to a new steady-state value after the light is quickly switched off. In this situation, the kinetics of the complex density can be understood by solving Eq. \ref{complexRateEq} with the initial condition $n_c(t=0) = n_0$, where $t=0$ is the time at which the light turns off, and $n_0$ is the suppressed complex density at and before the turn-off. The solution is given by
\begin{equation} \label{complexGrowth}
n_c(t) = \gamma\tau_c\left[1 - \left( 1 - \frac{n_0}{\gamma\tau_c} \right)\textrm{e}^{-t/\tau_c}\right],
\end{equation}
where the $1/e$ saturation time is the complex lifetime $\tau_c$. Note that Eq. \ref{complexGrowth} is exact if the light turns off instantaneously, and is a good approximation if the characteristic turn-off time $t_{\textrm{off}} \ll \tau_c$. To minimize $t_{\textrm{off}}$, we introduce a ``kill'' beam whose wavelength and optical path are identical to the ``H'' beam (Fig. \ref{figSchematic}(a)), but whose timing and intensity are independently controlled. This ``kill'' beam has a 90 - 10\% fall time of 30 ns, compared to $> 200$ ns for the ``H'' and ``V'' beams.

As the reaction proceeds, the intensities of the ODT beams are modulated with 100\% depth at 7 kHz. During the dark phase of the modulation, the ``kill'' beam is pulsed on for 10 $\mu s$ with an intensity of 12.7 kW/cm$^2$, and the UV pulse probes the complex population at various delay times around the off edge of the ``kill'' pulse (Fig. \ref{figComplexLifetime}(a)). Here we define $t = 0$ as the time when the UV pulse is 27 ns delayed with respect to the midpoint of the ``kill'' pulse turn-off. The K$_2$Rb$_2^+$ counts are plotted against the UV delay in Fig. \ref{figComplexLifetime}(b) and fitted using the function $N_{\textrm{K}_2\textrm{Rb}_2^+} = A\left[1 - (1-N_0/A)\textrm{e}^{-(t - t_0)/\tau_c}\right]$, derived from Eq. \ref{complexGrowth}, with $A$, $t_0$, and $\tau_c$ as free parameters. Here, $t_0$ is introduced to account for a systematic timing uncertainty; $N_0$ is fixed to the ion counts measured at a UV delay of $-1~\mu\textrm{s}$, when the ``kill'' pulse is still on, which serves as a proxy for $n_0$. The complex lifetime given by the fit is $360 \pm 30$ ns.

Having measured both $\tau_c$ and $B_{1,2} = \beta_{1,2} \tau_c$, we can determine the first- and second-order rate constants, $\beta_1$ and $\beta_2$, for complex excitation by 1064 nm light. For $B_1 = 0.15 \pm 0.03~\textrm{cm}^2/\textrm{kW}$ and $B_2 = 0.050 \pm 0.004~(\textrm{cm}^2/\textrm{kW})^2$ extracted from the fit to the data in Fig. \ref{figODTEffect}, we obtain $\beta_1 = 0.42 \pm 0.09 ~\mu\textrm{s}^{-1}/(\textrm{kW/cm}^2)$ and $\beta_2 = 0.14 \pm 0.02 ~\mu\textrm{s}^{-1}/\left(\textrm{kW/cm}^2\right)^2$.

\section{Theory and comparison}

Finally, we perform theoretical calculations of both the lifetime and the photo-excitation rate of the complex and compare them to the measurements.
We assume the complex explores the reaction phase space ergodically before dissociating into products/reactants and  estimate the lifetime using RRKM theory \cite{levine2009molecular}, $\tau_c = 2\pi\hbar \rho_c/\mathcal{N}$. Here, $\rho_c$ is the density of states (DOS) of the complex, and $\mathcal{N}$ is the number of open product channels via which the complex can dissociate. We also assume that, over the course of the reaction, the total angular momentum of the system ($J = 1$) \footnote{The reactant KRb molecules, which are prepared in a single hyperfine quantum state, are identical fermions. Therefore, their collisions at ultralow temperatures are restricted to p-wave, with angular momentum $L = 1$. Since the KRb molecules are in the rotational ground state ($N = 0$), the total angular momentum (\textbf{J} = \textbf{N} + \textbf{L}) for the KRb + KRb system is $J = 1$} is conserved, and the spins of the nuclei remain unchanged. $\mathcal{N}$ is counted using known values for the exothermicity of the reaction \cite{ospelkaus2010quantum} and the rotational constants of K$_2$ \cite{huber2013molecular} and Rb$_2$ \cite{amiot1990laser}. Computing $\rho_c$ requires knowledge of the ground potential energy surface (PES) for the KRb + KRb system, which we calculate in full dimensionality using the {\sc Molpro} package \cite{MOLPRO_brief} (see methods). A one-dimensional cut of the PES is shown in Fig. \ref{figTheory}(a). Using this PES and following the quasi-classical method developed in Ref. \cite{christianen2019quasiclassical}, we calculate $\rho_c$ to be $2.6 \pm 0.8~\mu$K$^{-1}$. The uncertainty is estimated by scaling the potential by $\pm 15\%$. The resulting RRKM lifetime is $170 \pm 60$ ns, which is remarkably similar to the  measured value.
While we have assumed that nuclear spins do not change during the reaction, the calculated lifetime is expected to be comparable even if they do, because any additional hyperfine channels will increase both $\rho_c$ and $\mathcal{N}$ by similar proportions, whereas the RRKM lifetime depends only on the ratio of the two.

To calculate the first-order photo-excitation rate constant, we compute excitation energies and transition dipole moments for the lowest twelve excited states. The single-photon excitation rate of the complex from the ground state, $i$, to a particular excited state, $f$, is computed as $\Gamma_{i\rightarrow f} = c^{-1} \langle b_{i\rightarrow f}(\omega) \rangle I$.
Here, $c$ is the speed of light, $I$ the optical intensity, $\omega$ the excitation frequency, and $b$ the Einstein B-coefficient for state $f$. Under the assumption that the complex ergodically explores the ground PES, $b$ can be calculated by averaging the squared transition dipole moment over the local DOS \cite{christianen2019photoinduced}. Contributions from different excited states are then summed to obtain the total excitation rate $\Gamma_l$. The rate constant, $\beta_1 = \Gamma_l/I$, is shown in Fig. \ref{figTheory}(b) as a function of the excitation photon energy. At 1064 nm this excitation rate is $\beta_1 = 0.4^{+0.4}_{-0.2} ~\mu\textrm{s}^{-1}/(\textrm{kW/cm}^2)$, which agrees well with the measured first-order excitation rate of $0.42 \pm 0.09 ~\mu\textrm{s}^{-1}/(\textrm{kW/cm}^2)$.
Calculation of the second-order rate constant, $\beta_2$, requires an understanding of higher-order excitation processes in the complex, which is beyond the scope of the current work.

\section{Conclusions and outlook}

Our identification of the photo-excited reactive pathway in an ultracold KRb gas not only brings new understanding to the origin of molecular loss in this much-studied system \cite{ospelkaus2010quantum,hu2019direct,de2019degenerate}, but also has important implications for the losses observed in other ultracold bialkali species whose ground state reactive pathways are energetically forbidden \cite{takekoshi2014ultracold,ye2018collisions, gregory2019sticky, park2015ultracold}. Because the electronic structures of these systems are comparable to KRb, the complexes formed from bimolecular collisions are also susceptible to excitations and losses induced by the trapping light, as suggested by Ref. \cite{christianen2019photoinduced}.
Avoiding this pathway may therefore facilitate the creation of molecular gases with higher phase space densities and the exploration of the many new scientific directions such systems promise~\cite{buchler2007strongly, levine2009molecular, baranov2012condensed, yao2018quantum}. From the perspective of controlling chemical reactions, the long-lived complex formed from ultracold molecular collisions provides a new handle to steer the reaction dynamics. Studying the decay pathways for the optically excited complex may allow for the direct control of reaction outcomes using light. By further combining quantum state readout of the reaction complexes and products, ultracold molecules offer the opportunity to investigate bond breaking and formation at the microscopic level.

\vspace{\fill}

\textit{Note added}. — During the preparation of this manuscript, we became aware of a related parallel experimental work \cite{Gregory2020loss}.

\clearpage

\section*{References}
\bibliography{refs}

\begin{thebibliography}{10}
\expandafter\ifx\csname url\endcsname\relax
  \def\url#1{\texttt{#1}}\fi
\expandafter\ifx\csname urlprefix\endcsname\relax\def\urlprefix{URL }\fi
\providecommand{\bibinfo}[2]{#2}
\providecommand{\eprint}[2][]{\url{#2}}

\bibitem{qiu2006observation}
\bibinfo{author}{Qiu, M.} \emph{et~al.}
\newblock \bibinfo{title}{Observation of {Feshbach} resonances in the {F +
  H$_2$} $ \rightarrow $ {HF + H} reaction}.
\newblock \emph{\bibinfo{journal}{Science}} \textbf{\bibinfo{volume}{311}},
  \bibinfo{pages}{1440--1443} (\bibinfo{year}{2006}).

\bibitem{pan2017direct}
\bibinfo{author}{Pan, H.}, \bibinfo{author}{Wang, F.},
  \bibinfo{author}{Czak{\'o}, G.} \& \bibinfo{author}{Liu, K.}
\newblock \bibinfo{title}{Direct mapping of the angle-dependent barrier to
  reaction for cl+ chd 3 using polarized scattering data}.
\newblock \emph{\bibinfo{journal}{Nature chemistry}}
  \textbf{\bibinfo{volume}{9}}, \bibinfo{pages}{1175} (\bibinfo{year}{2017}).

\bibitem{puri2019reaction}
\bibinfo{author}{Puri, P.} \emph{et~al.}
\newblock \bibinfo{title}{Reaction blockading in a reaction between an excited
  atom and a charged molecule at low collision energy}.
\newblock \emph{\bibinfo{journal}{Nature chemistry}}
  \textbf{\bibinfo{volume}{11}}, \bibinfo{pages}{615} (\bibinfo{year}{2019}).

\bibitem{levine2009molecular}
\bibinfo{author}{Levine, R.~D.}
\newblock \emph{\bibinfo{title}{Molecular reaction dynamics}}
  (\bibinfo{publisher}{Cambridge University Press}, \bibinfo{year}{2009}).

\bibitem{ospelkaus2010quantum}
\bibinfo{author}{Ospelkaus, S.} \emph{et~al.}
\newblock \bibinfo{title}{Quantum-state controlled chemical reactions of
  ultracold potassium-rubidium molecules}.
\newblock \emph{\bibinfo{journal}{Science}} \textbf{\bibinfo{volume}{327}},
  \bibinfo{pages}{853--857} (\bibinfo{year}{2010}).

\bibitem{takekoshi2014ultracold}
\bibinfo{author}{Takekoshi, T.} \emph{et~al.}
\newblock \bibinfo{title}{Ultracold dense samples of dipolar {RbCs} molecules
  in the rovibrational and hyperfine ground state}.
\newblock \emph{\bibinfo{journal}{Physical Review Letters}}
  \textbf{\bibinfo{volume}{113}}, \bibinfo{pages}{205301}
  (\bibinfo{year}{2014}).

\bibitem{ye2018collisions}
\bibinfo{author}{Ye, X.}, \bibinfo{author}{Guo, M.},
  \bibinfo{author}{Gonz{\'a}lez-Mart{\'\i}nez, M.~L.},
  \bibinfo{author}{Qu{\'e}m{\'e}ner, G.} \& \bibinfo{author}{Wang, D.}
\newblock \bibinfo{title}{Collisions of ultracold $^{23}${Na}$^{87}${Rb}
  molecules with controlled chemical reactivities}.
\newblock \emph{\bibinfo{journal}{Science advances}}
  \textbf{\bibinfo{volume}{4}}, \bibinfo{pages}{eaaq0083}
  (\bibinfo{year}{2018}).

\bibitem{gregory2019sticky}
\bibinfo{author}{Gregory, P.~D.} \emph{et~al.}
\newblock \bibinfo{title}{Sticky collisions of ultracold rbcs molecules}.
\newblock \emph{\bibinfo{journal}{Nature communications}}
  \textbf{\bibinfo{volume}{10}}, \bibinfo{pages}{1--7} (\bibinfo{year}{2019}).

\bibitem{park2015ultracold}
\bibinfo{author}{Park, J.~W.}, \bibinfo{author}{Will, S.~A.} \&
  \bibinfo{author}{Zwierlein, M.~W.}
\newblock \bibinfo{title}{Ultracold dipolar gas of fermionic na 23 k 40
  molecules in their absolute ground state}.
\newblock \emph{\bibinfo{journal}{Physical Review Letters}}
  \textbf{\bibinfo{volume}{114}}, \bibinfo{pages}{205302}
  (\bibinfo{year}{2015}).

\bibitem{ni2010dipolar}
\bibinfo{author}{Ni, K.-K.} \emph{et~al.}
\newblock \bibinfo{title}{Dipolar collisions of polar molecules in the quantum
  regime}.
\newblock \emph{\bibinfo{journal}{Nature}} \textbf{\bibinfo{volume}{464}},
  \bibinfo{pages}{1324} (\bibinfo{year}{2010}).

\bibitem{de2011controlling}
\bibinfo{author}{De~Miranda, M.} \emph{et~al.}
\newblock \bibinfo{title}{Controlling the quantum stereodynamics of ultracold
  bimolecular reactions}.
\newblock \emph{\bibinfo{journal}{Nature Physics}}
  \textbf{\bibinfo{volume}{7}}, \bibinfo{pages}{502} (\bibinfo{year}{2011}).

\bibitem{hall2012millikelvin}
\bibinfo{author}{Hall, F.~H.} \& \bibinfo{author}{Willitsch, S.}
\newblock \bibinfo{title}{Millikelvin reactive collisions between
  sympathetically cooled molecular ions and laser-cooled atoms in an ion-atom
  hybrid trap}.
\newblock \emph{\bibinfo{journal}{Physical Review Letters}}
  \textbf{\bibinfo{volume}{109}}, \bibinfo{pages}{233202}
  (\bibinfo{year}{2012}).

\bibitem{mcdonald2016photodissociation}
\bibinfo{author}{McDonald, M.} \emph{et~al.}
\newblock \bibinfo{title}{Photodissociation of ultracold diatomic strontium
  molecules with quantum state control}.
\newblock \emph{\bibinfo{journal}{Nature}} \textbf{\bibinfo{volume}{535}},
  \bibinfo{pages}{122} (\bibinfo{year}{2016}).

\bibitem{klein2017directly}
\bibinfo{author}{Klein, A.} \emph{et~al.}
\newblock \bibinfo{title}{Directly probing anisotropy in atom--molecule
  collisions through quantum scattering resonances}.
\newblock \emph{\bibinfo{journal}{Nature Physics}}
  \textbf{\bibinfo{volume}{13}}, \bibinfo{pages}{35} (\bibinfo{year}{2017}).

\bibitem{yang2019observation}
\bibinfo{author}{Yang, H.} \emph{et~al.}
\newblock \bibinfo{title}{Observation of magnetically tunable feshbach
  resonances in ultracold 23na40k+ 40k collisions}.
\newblock \emph{\bibinfo{journal}{Science}} \textbf{\bibinfo{volume}{363}},
  \bibinfo{pages}{261--264} (\bibinfo{year}{2019}).

\bibitem{rui2017controlled}
\bibinfo{author}{Rui, J.} \emph{et~al.}
\newblock \bibinfo{title}{Controlled state-to-state atom-exchange reaction in
  an ultracold atom--dimer mixture}.
\newblock \emph{\bibinfo{journal}{Nature Physics}}
  \textbf{\bibinfo{volume}{13}}, \bibinfo{pages}{699} (\bibinfo{year}{2017}).

\bibitem{hoffmann2018reaction}
\bibinfo{author}{Hoffmann, D.~K.}, \bibinfo{author}{Paintner, T.},
  \bibinfo{author}{Limmer, W.}, \bibinfo{author}{Petrov, D.~S.} \&
  \bibinfo{author}{Denschlag, J.~H.}
\newblock \bibinfo{title}{Reaction kinetics of ultracold molecule-molecule
  collisions}.
\newblock \emph{\bibinfo{journal}{Nature communications}}
  \textbf{\bibinfo{volume}{9}}, \bibinfo{pages}{5244} (\bibinfo{year}{2018}).

\bibitem{hu2019direct}
\bibinfo{author}{Hu, M.-G.} \emph{et~al.}
\newblock \bibinfo{title}{Direct observation of bimolecular reactions of
  ultracold krb molecules}.
\newblock \emph{\bibinfo{journal}{Science}} \textbf{\bibinfo{volume}{366}},
  \bibinfo{pages}{1111--1115} (\bibinfo{year}{2019}).

\bibitem{santos2000bose}
\bibinfo{author}{Santos, L.}, \bibinfo{author}{Shlyapnikov, G.},
  \bibinfo{author}{Zoller, P.} \& \bibinfo{author}{Lewenstein, M.}
\newblock \bibinfo{title}{Bose-einstein condensation in trapped dipolar gases}.
\newblock \emph{\bibinfo{journal}{Physical Review Letters}}
  \textbf{\bibinfo{volume}{85}}, \bibinfo{pages}{1791} (\bibinfo{year}{2000}).

\bibitem{buchler2007strongly}
\bibinfo{author}{B{\"u}chler, H.~P.} \emph{et~al.}
\newblock \bibinfo{title}{Strongly correlated 2d quantum phases with cold polar
  molecules: controlling the shape of the interaction potential}.
\newblock \emph{\bibinfo{journal}{Physical Review Letters}}
  \textbf{\bibinfo{volume}{98}}, \bibinfo{pages}{060404}
  (\bibinfo{year}{2007}).

\bibitem{levinsen2011topological}
\bibinfo{author}{Levinsen, J.}, \bibinfo{author}{Cooper, N.~R.} \&
  \bibinfo{author}{Shlyapnikov, G.~V.}
\newblock \bibinfo{title}{Topological p x+ ip y superfluid phase of fermionic
  polar molecules}.
\newblock \emph{\bibinfo{journal}{Physical Review A}}
  \textbf{\bibinfo{volume}{84}}, \bibinfo{pages}{013603}
  (\bibinfo{year}{2011}).

\bibitem{baranov2012condensed}
\bibinfo{author}{Baranov, M.~A.}, \bibinfo{author}{Dalmonte, M.},
  \bibinfo{author}{Pupillo, G.} \& \bibinfo{author}{Zoller, P.}
\newblock \bibinfo{title}{Condensed matter theory of dipolar quantum gases}.
\newblock \emph{\bibinfo{journal}{Chemical Reviews}}
  \textbf{\bibinfo{volume}{112}}, \bibinfo{pages}{5012--5061}
  (\bibinfo{year}{2012}).

\bibitem{christianen2019photoinduced}
\bibinfo{author}{Christianen, A.}, \bibinfo{author}{Zwierlein, M.~W.},
  \bibinfo{author}{Groenenboom, G.~C.} \& \bibinfo{author}{Karman, T.}
\newblock \bibinfo{title}{Photoinduced two-body loss of ultracold molecules}.
\newblock \emph{\bibinfo{journal}{Physical Review Letters}}
  \textbf{\bibinfo{volume}{123}}, \bibinfo{pages}{123402}
  (\bibinfo{year}{2019}).

\bibitem{mayle2013scattering}
\bibinfo{author}{Mayle, M.}, \bibinfo{author}{Qu{\'e}m{\'e}ner, G.},
  \bibinfo{author}{Ruzic, B.~P.} \& \bibinfo{author}{Bohn, J.~L.}
\newblock \bibinfo{title}{Scattering of ultracold molecules in the highly
  resonant regime}.
\newblock \emph{\bibinfo{journal}{Physical Review A}}
  \textbf{\bibinfo{volume}{87}}, \bibinfo{pages}{012709}
  (\bibinfo{year}{2013}).

\bibitem{christianen2019quasiclassical}
\bibinfo{author}{Christianen, A.}, \bibinfo{author}{Karman, T.} \&
  \bibinfo{author}{Groenenboom, G.~C.}
\newblock \bibinfo{title}{Quasiclassical method for calculating the density of
  states of ultracold collision complexes}.
\newblock \emph{\bibinfo{journal}{Physical Review A}}
  \textbf{\bibinfo{volume}{100}}, \bibinfo{pages}{032708}
  (\bibinfo{year}{2019}).

\bibitem{liu2020probing}
\bibinfo{author}{Liu, Y.}, \bibinfo{author}{Grimes, D.~D.},
  \bibinfo{author}{Hu, M.-G.} \& \bibinfo{author}{Ni, K.-K.}
\newblock \bibinfo{title}{Probing ultracold chemistry using ion spectrometry}.
\newblock \emph{\bibinfo{journal}{Physical Chemistry Chemical Physics}}
  (\bibinfo{year}{2020}).

\bibitem{vadla2006comparison}
\bibinfo{author}{Vadla, C.} \emph{et~al.}
\newblock \bibinfo{title}{Comparison of theoretical and experimental red and
  near infrared absorption spectra in overheated potassium vapour}.
\newblock \emph{\bibinfo{journal}{The European Physical Journal D-Atomic,
  Molecular, Optical and Plasma Physics}} \textbf{\bibinfo{volume}{37}},
  \bibinfo{pages}{37--49} (\bibinfo{year}{2006}).

\bibitem{edvardsson2003calculation}
\bibinfo{author}{Edvardsson, D.}, \bibinfo{author}{Lunell, S.} \&
  \bibinfo{author}{Marian, C.~M.}
\newblock \bibinfo{title}{Calculation of potential energy curves for {Rb}$_2$
  including relativistic effects}.
\newblock \emph{\bibinfo{journal}{Molecular Physics}}
  \textbf{\bibinfo{volume}{101}}, \bibinfo{pages}{2381--2389}
  (\bibinfo{year}{2003}).

\bibitem{sato2001photodissociation}
\bibinfo{author}{Sato, H.}
\newblock \bibinfo{title}{Photodissociation of simple molecules in the gas
  phase}.
\newblock \emph{\bibinfo{journal}{Chemical Reviews}}
  \textbf{\bibinfo{volume}{101}}, \bibinfo{pages}{2687--2726}
  (\bibinfo{year}{2001}).

\bibitem{noll1998bimolecular}
\bibinfo{author}{Noll, R.~J.}, \bibinfo{author}{Yi, S.~S.} \&
  \bibinfo{author}{Weisshaar, J.~C.}
\newblock \bibinfo{title}{Bimolecular ni+ (2d5/2)+ c3h8 reaction dynamics in
  real time}.
\newblock \emph{\bibinfo{journal}{The Journal of Physical Chemistry A}}
  \textbf{\bibinfo{volume}{102}}, \bibinfo{pages}{386--394}
  (\bibinfo{year}{1998}).

\bibitem{scherer1990real}
\bibinfo{author}{Scherer, N.}, \bibinfo{author}{Sipes, C.},
  \bibinfo{author}{Bernstein, R.} \& \bibinfo{author}{Zewail, A.}
\newblock \bibinfo{title}{Real-time clocking of bimolecular reactions:
  Application to h+ co2}.
\newblock \emph{\bibinfo{journal}{The Journal of Chemical Physics}}
  \textbf{\bibinfo{volume}{92}}, \bibinfo{pages}{5239--5259}
  (\bibinfo{year}{1990}).

\bibitem{Note1}
\bibinfo{note}{The reactant KRb molecules, which are prepared in a single
  hyperfine quantum state, are identical fermions. Therefore, their collisions
  at ultralow temperatures are restricted to p-wave, with angular momentum $L =
  1$. Since the KRb molecules are in the rotational ground state ($N = 0$), the
  total angular momentum (\protect \textbf {J} = \protect \textbf {N} +
  \protect \textbf {L}) for the KRb + KRb system is $J = 1$}.

\bibitem{huber2013molecular}
\bibinfo{author}{Huber, K.-P.}
\newblock \emph{\bibinfo{title}{Molecular spectra and molecular structure: IV.
  Constants of diatomic molecules}} (\bibinfo{publisher}{Springer Science \&
  Business Media}, \bibinfo{year}{2013}).

\bibitem{amiot1990laser}
\bibinfo{author}{Amiot, C.}
\newblock \bibinfo{title}{Laser-induced fluorescence of rb2: The (1) 1$\sigma$+
  g (x),(2) 1$\sigma$+ g,(1) 1$\pi$ u (b),(1) 1$\pi$ g, and (2) 1$\pi$ u (c)
  electronic states}.
\newblock \emph{\bibinfo{journal}{The Journal of chemical physics}}
  \textbf{\bibinfo{volume}{93}}, \bibinfo{pages}{8591--8604}
  (\bibinfo{year}{1990}).

\bibitem{MOLPRO_brief}
\bibinfo{author}{Werner, H.-J.} \emph{et~al.}
\newblock \bibinfo{title}{Molpro, version 2019.2, a package of ab initio
  programs} (\bibinfo{year}{2019}).
\newblock \bibinfo{note}{See www.molpro.net}.

\bibitem{de2019degenerate}
\bibinfo{author}{De~Marco, L.} \emph{et~al.}
\newblock \bibinfo{title}{A degenerate fermi gas of polar molecules}.
\newblock \emph{\bibinfo{journal}{Science}} \textbf{\bibinfo{volume}{363}},
  \bibinfo{pages}{853--856} (\bibinfo{year}{2019}).

\bibitem{yao2018quantum}
\bibinfo{author}{Yao, N.~Y.}, \bibinfo{author}{Zaletel, M.~P.},
  \bibinfo{author}{Stamper-Kurn, D.~M.} \& \bibinfo{author}{Vishwanath, A.}
\newblock \bibinfo{title}{A quantum dipolar spin liquid}.
\newblock \emph{\bibinfo{journal}{Nature Physics}}
  \textbf{\bibinfo{volume}{14}}, \bibinfo{pages}{405--410}
  (\bibinfo{year}{2018}).

\bibitem{Gregory2020loss}
\bibinfo{author}{Gregory, P.~D.}, \bibinfo{author}{Blackmore, J.~A.},
  \bibinfo{author}{Bromley, S.~L.} \& \bibinfo{author}{Cornish, S.~L.}
\newblock \bibinfo{title}{Loss of ultracold $^{87}${Rb}$^{133}${Cs} molecules
  via optical excitation of long-lived two-body collision complexes}.
\newblock \emph{\bibinfo{journal}{arXiv: 2002.04431}}  (\bibinfo{year}{2020}).

\bibitem{fuentealba1982proper}
\bibinfo{author}{Fuentealba, P.}, \bibinfo{author}{Preuss, H.},
  \bibinfo{author}{Stoll, H.} \& \bibinfo{author}{Von~Szentp{\'a}ly, L.}
\newblock \bibinfo{title}{A proper account of core-polarization with
  pseudopotentials: single valence-electron alkali compounds}.
\newblock \emph{\bibinfo{journal}{Chemical Physics Letters}}
  \textbf{\bibinfo{volume}{89}}, \bibinfo{pages}{418--422}
  (\bibinfo{year}{1982}).

\bibitem{christianen2019six}
\bibinfo{author}{Christianen, A.}, \bibinfo{author}{Karman, T.},
  \bibinfo{author}{Vargas-Hern{\'a}ndez, R.~A.}, \bibinfo{author}{Groenenboom,
  G.~C.} \& \bibinfo{author}{Krems, R.~V.}
\newblock \bibinfo{title}{Six-dimensional potential energy surface for nak--nak
  collisions: Gaussian process representation with correct asymptotic form}.
\newblock \emph{\bibinfo{journal}{The Journal of chemical physics}}
  \textbf{\bibinfo{volume}{150}}, \bibinfo{pages}{064106}
  (\bibinfo{year}{2019}).

\bibitem{byrd2010structure}
\bibinfo{author}{Byrd, J.~N.}, \bibinfo{author}{Montgomery~Jr, J.~A.} \&
  \bibinfo{author}{C{\^o}t{\'e}, R.}
\newblock \bibinfo{title}{Structure and thermochemistry of {K$_2$Rb, KRb$_2$,
  and K$_2$Rb$_2$}}.
\newblock \emph{\bibinfo{journal}{Physical Review A}}
  \textbf{\bibinfo{volume}{82}}, \bibinfo{pages}{010502}
  (\bibinfo{year}{2010}).

\end{thebibliography}
\bibliographystyle{naturemag}

\clearpage

\section*{Methods}

\textbf{Experimental setup} The ODT for the KRb molecules is formed by the crossing of two Gaussian beams, ``H'' and ``V'', with $1/e^2$ waist diameters of 60 and 200 $\mu$m, respectively, at a 70$^\circ$ angle (Fig. \ref{figSchematic}(a)). At a total optical intensity of 11.3 $\textrm{kW}/\textrm{cm}^2$, this trap configuration results in a cigar-shaped cloud of KRb molecules with 2$\sigma$ Gaussian widths of 6, 6, and 28 $\mu\textrm{m}$ along its three principal axes. Thus, the ODT intensity varies by less than $4 \%$ over these widths and is considered to be  constant across the sample. All ODT beams are derived from a single 1064 nm laser source with a spectral width of 1 kHz.

Weak static electric (17 V/cm) and magnetic (5 G) fields are applied during the experiment, for the respective purpose of extracting ions for TOF-MS and maintaining nulcear spin quantization for the KRb molecules. The direction of the electric field is normal to the plane containing the ODT beams, whereas the magnetic field is anti-parallel to the gravity direction.

\textbf{Quantum states of the reactants and the products} The reactant KRb molecules are prepared in the $ |m_I^\text{K} = -4, m_I^\text{Rb} = 1/2 \rangle $ hyperfine state of the lowest rovibronic state, $ |X^1 \Sigma^+, v = 0, N = 0 \rangle$, where $m_I$, $v$, and $N$ are the nuclear magnetic spin, molecular vibration, and molecular rotation quantum numbers, respectively. The product K$_2$ and Rb$_2$ molecules emerge from reactions in their respective vibrational ground states, but with rotational excitations of up to $N_{\textrm{K}_2} = 11$ and $N_{\textrm{Rb}_2} = 19$.

\textbf{Photo-ionization of the products and the complex} As the reaction proceeds, the products, K$_2$ and Rb$_2$, escape from the ODT due to translational energy gained from the exothermicity, and establish steady-state density distributions, $n_{\textrm{K}_2}$ and $n_{\textrm{Rb}_2}$, around the KRb cloud. Since products are generated at the same rate that the K$_2$Rb$_2^*$ complex dissociates, $n_{\textrm{K}_2}$ and $n_{\textrm{Rb}_2}$ are both proportional to the complex density $n_c$. The photon energy at the wavelength used to ionize products (305 nm) is above the ionization threshold of KRb. In order to selectively ionize the products, and not the reactants inside the ODT, the pulsed UV laser is shaped into a ring-shaped beam profile centered around the KRb cloud, with a diameter of 0.45 mm and a $1\sigma$ Gaussian width of 5.4 $\mu$m. The complexes, on the other hand, reside within the ODT due to their negligible lab-frame translational energy and their transient nature. To photo-ionize them, the UV laser is shaped into a Gaussian beam with a $1/e^2$ waist diameter of 300 $\mu$m that overlaps the KRb cloud. Fortunately, the photon energy at the wavelength used here (354.77 nm) is below the ionization threshold of KRb and does not result in significant depletion of the KRb molecules.

\textbf{Ion counts as proxies for product and complex densities}
While the data presented in Fig. \ref{figODTEffect} are in the form of ion counts, $N_{\textrm{K}_2^+}$, $N_{\textrm{Rb}_2^+}$, and $N_{\textrm{K}_2\textrm{Rb}_2^+}$, the modeling of these data are based on the instantaneous density of the K$_2$Rb$_2^*$ complex, $n_c$. Here, we establish the proportionality between the detected ion counts and the complex density, starting with a derivation of the relationship between $N_{\textrm{K}_2\textrm{Rb}_2^+}$ (Eq. \ref{IonSteadyState}) and $n_c$ (Eq. \ref{complexDensitySteadyState}).
The pre-factor $\gamma = -\dot{n}_r/2$ in Eq. \ref{complexDensitySteadyState} describes the decay of the complex density within each experimental cycle arising from the decay of the KRb population, which has a half-life of hundreds of milliseconds. Consider a single ionization UV pulse that probes the system at a time $t$ after the creation of the KRb sample. Since the UV pulse is a few nanoseconds in duration, it maps the complex density at that time onto a detected ion count
\begin{equation}
N_{\textrm{K}_2\textrm{Rb}_2^+}^{{\textrm{single}}}(t) = P_{\textrm{ion}} \eta_{\textrm{det}} V_{\textrm{trap}} \gamma(t) \frac{1}{\tau_c^{-1} + \Gamma_l}, 
\end{equation}
where $P_{\textrm{ion}}$ is the ionization probability, $\eta_{\textrm{det}}$ is the ion detection efficiency, and $V_{\textrm{trap}}$ is the effective volume of the trap. In each experimental cycle, a newly-prepared KRb sample is probed by a total of 7000 UV pulses over a duration of 1 s. The total ion count per cycle is therefore
\begin{equation}
N_{\textrm{K}_2\textrm{Rb}_2^+} = \left[P_{\textrm{ion}} \eta_{\textrm{det}} V_{\textrm{trap}} \sum_{m = 1}^{7000} \gamma(mT)\right] \frac{1}{\tau_c^{-1} + \Gamma_l},
\end{equation}
where $T = 143~\mu$s is the period of the UV pulses. Note that the bracketed pre-factor is a constant, which we call $A$ in Eq. \ref{IonSteadyState}. This shows that $N_{\textrm{K}_2\textrm{Rb}_2^+}$ is indeed proportional to $n_c$. Following analogous arguments, it can be shown that the product ion counts, $N_{\textrm{K}_2^+}$ and $N_{\textrm{Rb}_2^+}$, are also proportional to the associated densities, $n_{\textrm{K}_2}$ and $n_{\textrm{Rb}_2}$, which are in turn proportional to $n_c$. The same logic applies with regard to the relationship between the time-dependent ion counts and the complex density for the complex lifetime measurement shown in Fig. \ref{figComplexLifetime}.

\textbf{Reaching a steady state following a change in the instantaneous ODT intensity} When the ODT intensity is switched from the $I'$-phase to the $I$-phase (Fig. \ref{figODTEffect} insets), the density of the complex evolves into a new steady-state value on a timescale equal to or less than the complex lifetime $\tau_c$, which we have measured to be $360 \pm 30$ ns.
The products, on the other hand, take longer a time to re-establish their steady-state density distribution around the KRb cloud, due to the finite speed with which they escape the ODT. This timescale, which depends on the diameter of the hollow UV beam (0.45 mm) and the typical speed of the products (a few to a few tens of meters per second), is empirically determined to be $\sim 100~\mu$s. Thus, for both the product and the complex measurements shown in Fig. \ref{figODTEffect}, there is a sufficient separation in time between the intensity switch and the UV pulse such that we probe the steady-state complex and product densities during the $I$-phase.

\setcounter{figure}{0}
\renewcommand{\thefigure}{\arabic{figure} Extended Data}

\begin{figure}[t!]
\centering
\includegraphics[width = 1.0\textwidth]{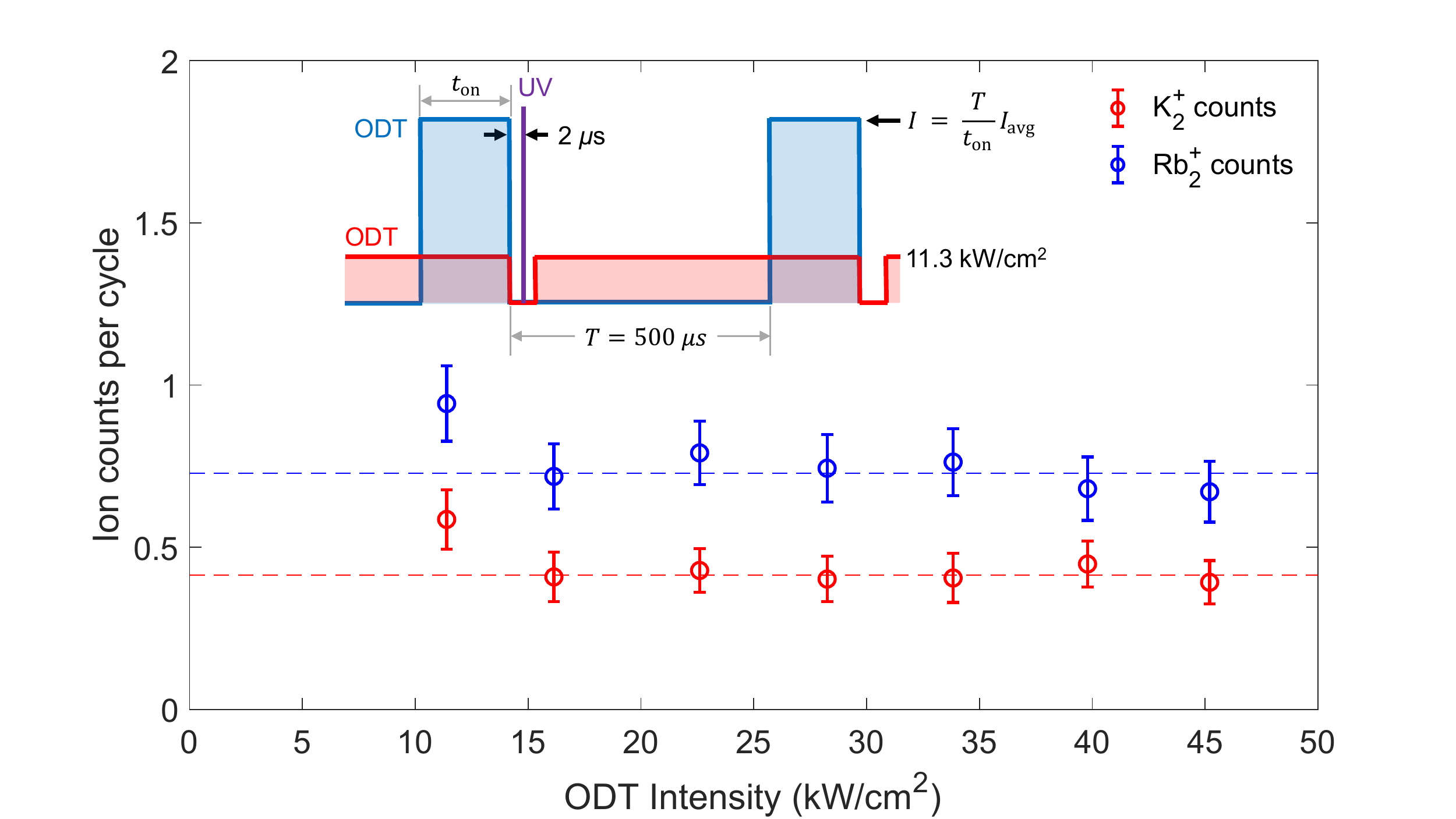}
\caption{\textbf{Continued formation of products at high ODT intensities} Steady-state K$_2^+$ (red circles) and Rb$_2^+$ (blue circles) ion counts at ODT light intensities in the 11.3 - 46.2 $\textrm{kW}/\textrm{cm}^2$ range, normalized by the number of experimental cycles ($\sim 80$ for each data point). The error bars represent shot noise.
The dashed lines indicate the levels to which the ion counts plateau, obtained by averaging, within each data set, the values of the points at ODT intensities larger than 15 $\textrm{kW}/\textrm{cm}^2$.
(Inset) Timing schemes of the ODT (red and blue) and the pulsed UV ionization laser (purple) used for the measurements presented here. The red (blue) trace corresponds to a high (low) duty cycle modulation of the ODT. The instantaneous ODT intensity, $I$, is inversely proportional to the duty cycle, while the time-averaged ODT intensity, $I_{\textrm{avg}}$ is constant for all measurements.}
\label{figE1}
\end{figure}

\textbf{Continued formation of products at high ODT intensities} We measure the dependence of the product ion counts on the ODT intensity in the 11.3 - 46.2 $\textrm{kW}/\textrm{cm}^2$ range, which partially overlaps with, but extends beyond the range probed in Fig. \ref{figODTEffect}(a). For these measurements, the intensity of the ODT is modulated with full contrast, but with a variable duty cycle $D = t_{\textrm{on}}/T$, where the ODT is on for a duration of $t_{\textrm{on}}$ within each cycle of period $T$, which is chosen to be 500 $\mu$s. The time-averaged ODT intensity, $I_{\textrm{avg}}$, is kept constant at 11.3 $\textrm{kW}/\textrm{cm}^2$. For a given duty cycle $D$, the intensity $I$ during the on-phase is $I = I_{\textrm{avg}}/D$. The UV ionization pulses then probe the products shortly (2 $\mu$s) after the ODT is turned off. As this time is much shorter than the time it takes to perturb the product density distribution at the location of the ionization ring ($\sim$100 $\mu$s), the detected product density accurately reflects that during the on-phase of the ODT. The normalized product ion counts at different ODT intensities are shown in Extended Data Figure 1. The inset shows the timing diagram for the intensities of the ODT and the pulsed UV ionization laser. We observe plateaus in the ion counts for $I \geq 15~\textrm{kW}/\textrm{cm}^2$, indicating continued product formation despite the high ODT intensity. This plateau justifies our choice of using the average value of the high intensity data points in Fig. \ref{figODTEffect}(a) to provide a baseline for the fitting.

\textbf{Theory methods}
To calculate the ground-state PES for the KRb + KRb system, we use polarizable effective core potentials \cite{fuentealba1982proper} and solve the electronic Schr\"odinger equation for the four valence electrons using multi-reference configuration interaction in a large one-electron basis set \cite{christianen2019six}. \emph{Ab initio} points are computed for approximately 2000 geometries, selected using Latin hypercube sampling, and fit using Gaussian Process regression \cite{christianen2019six}.
The minimum energy structures agree with Byrd \emph{et al.} \cite{byrd2010structure}, and the corresponding binding energies differ by less than 15~\%, which we take to be the uncertainty of the energy calculation.

Excitation energies and transition dipole moments are calculated at the complete active space self-consistent field level, using methods otherwise similar to the ground state calculation.
The uncertainty in the calculated excitation rate is limited by the poor fit of the transition dipole surfaces,
caused by crossings of adiabatic excited states at which the transition dipole functions are not analytic.
Therefore, we also perform calculations where we fit the total squared transition dipole moment, which is unaffected by crossings between excited states, and distribute this squared dipole moment equally over all excited states.
The resulting excitation rate is smaller by a factor of two (Fig. \ref{figTheory}(b)), which we take to be indicative of the uncertainty due to the dipole surfaces.

\clearpage

\textbf{Data availability} The data that support the findings of this study are available from the corresponding author upon reasonable request.

\bigskip

\textbf{Acknowledgements} We thank Lingbang Zhu for experimental assistance. This work is supported by DOE YIP and the David and Lucile Packard Foundation. M.A.N. is supported by a HQI postdoctoral fellowship. T.K. is supported by NWO Rubicon Grant No. 019.172EN.007 and the NSF through ITAMP. H.G. acknowledges a MURI grant from ARO (W911NF-19-1-0283).

\bigskip

\textbf{Author contributions} The experimental work and data analysis were carried out by Y.L., M.-G.H., M.A.N., D.D.G., and K.-K.N. Theoretical calculations were done by T.K. and H.G.. All authors contributed to interpreting the results and writing the manuscript.

\bigskip

\textbf{Competing interests} The authors declare no competing financial interests.

\bigskip

\textbf{Corresponding author} Correspondence to Yu Liu (e-mail: yuliu@g.harvard.edu)

\clearpage

\end{document}